\documentclass[aps,prb,twocolumn,showpacs,superscriptaddress]{revtex4}

\usepackage{graphicx}
\usepackage{amsmath}
\usepackage{amssymb}

\begin{document}

\title{Charge order driven by Fermi-arc instability and its connection with pseudogap in cuprate superconductors}

\author{Shiping Feng\footnote{Corresponding author. E-mail: spfeng@bnu.edu.cn} and Deheng Gao}

\affiliation{Department of Physics, Beijing Normal University, Beijing 100875, China}

\author{Huaisong Zhao}

\affiliation{College of Physics, Qingdao University, Qingdao 266071, China}

\begin{abstract}
The recently discovered charge order is a generic feature of cuprate superconductors, however, its microscopic origin remains debated. Within the framework of the fermion-spin theory, the nature of charge order in the pseudogap phase and its evolution with doping are studied by taking into account the electron self-energy (then the pseudogap) effect. It is shown that the antinodal region of the electron Fermi surface is suppressed by the electron self-energy, and then the low-energy electron excitations occupy the disconnected Fermi arcs located around the nodal region. In particular, the charge-order state is driven by the Fermi-arc instability, with a characteristic wave vector corresponding to the hot spots of the Fermi arcs rather than the antinodal nesting vector. Moreover, although the Fermi arc increases its length as a function of doping, the charge-order wave vector reduces almost linearity with the increase of doping. The theory also indicates that the Fermi arc, charge order, and pseudogap in cuprate superconductors are intimately related each other, and all of them emanates from the electron self-energy due to the interaction between electrons by the exchange of spin excitations.
\end{abstract}


\maketitle

\section{Introduction}

After intensive investigations over several years, a large body of experimental evidences has suggested the presence of charge order, or equivalently charge-density-wave, in the pseudogap phase of cuprate superconductors \cite{Comin15,Wu11,Chang12,Ghiringhelli12,Ghiringhelli14,Comin14,Neto14,Hashimoto15,Comin15a}. In particular, it has been argued that charge order may be a natural consequence of an electron Fermi surface (EFS) instability competing with superconductivity \cite{Comin15,Ghiringhelli12,Comin14,Neto14,Hashimoto15}. In the early days of angle-resolved photoemission spectroscopy (ARPES) measurements on cuprate superconductors in the normal-state, it is shown that EFS forms a continuous contour in momentum space in the entire doping range, and then the underlying EFS satisfies Luttinger's theorem \cite{Takahashi89,Campuzano90,Olson90,Marshall96,Loeser96,Ding97}. Later, with the improvements in the resolution of ARPES experiments, the observed data indicate that in the underdoped and optimally doped regimes, although the normal-state of cuprate superconductors is metallic, EFS around the antinodal region is suppressed, leading to that the low-energy electron excitations occupy disconnected segments called as the Fermi arcs located at the nodal region of the Brillouin zone \cite{Norman98,Kanigel06,Yoshida06,Tanaka06,Kanigel07,Nakayama09,Yoshida09,Meng11,Ideta12,Kondo13}, however, the underlying EFS determined from the low-energy spectral weight still fulfills Luttinger's theorem \cite{Yoshida06}. Very recently, some ARPES experimental results \cite{Comin15,Comin14,Hashimoto15} show that charge order emerges just below the pseudogap crossover temperature $T^{*}$, with a characteristic charge-order wave vector that rules out simple antinodal nesting in the single-particle limit but matches well with a phenomenological model of a many-body instability of the Fermi arcs. Furthermore, it is shown that the charge-order state is particularly obvious in the underdoped regime, and then the magnitude of the charge-order wave vector $Q_{\rm CO}$ smoothly decreases upon increasing doping \cite{Comin14}. These ARPES experimental results \cite{Comin15,Comin14,Hashimoto15} on the other hand indicate that there is an intrinsic link between the low-energy electronic structure and charge order in the pseudogap phase of cuprate superconductors.

Although the experimental observations from different measurement techniques \cite{Comin15,Wu11,Chang12,Ghiringhelli12,Ghiringhelli14,Comin14,Neto14,Hashimoto15,Comin15a} have confirmed that charge order is a generic feature of cuprate superconductors, its full understanding is still a challenging issue. Theoretically, the possible relationship between the pseudogap and charge order has been extensively studied \cite{Lee14,Harrison14,Sachdev13,Meier13,Atkinson15}. On the one hand, it has been postulated that charge order is an instability of the Fermi arcs, in which the Fermi arcs themselves result from an EFS instability around the antinodal region that is distinct from charge order \cite{Lee14}. On the other hand, it has been suggested that the charge-order wave vectors spanning the tips of the Fermi arcs are a direct signature of the pseudogap formation due to charge order, rather than being suggestive of pre-existing Fermi arcs that are unstable to charge order \cite{Harrison14,Sachdev13,Meier13}. In particular, within a three-band model, it has been shown that the antiferromagnetic (AF) correlations lead to a pseudogap-like reconstruction of EFS, then charge-order emerges from this pseudogap state, and is not the primary source of the pseudogap \cite{Atkinson15}. However, up to now, the nature of charge order has not been discussed starting from a microscopic superconducting (SC) mechanism, and no explicit calculations of the doping dependence of the charge-order wave vector $Q_{\rm CO}$ has been made so far. Superconductivity in cuprate superconductors is realized when charge carriers are doped into a parent Mott insulating state. This Mott insulating state emergences to be due to the strong electron correlation \cite{Phillips10}, while the strong electron correlation manifests itself by the no-double electron occupancy constraint in the system. To the best of our knowledge, there are two complementary approaches that have been used popularly to treat the no-double electron occupancy constraint: (a) Within the numerical techniques \cite{Dagotto94}, the no-double electron occupancy constraint can be treated exactly at {\it zero temperature}. In particular, a number of the numerical simulations has demonstrated, at various levels of rigour, that EFS of cuprate superconductors in the normal-state is consistent with a large EFS \cite{Dagotto94,Stephan91,Dagotto92,Moreo90} with the area that is given by $1-\delta$ as expected from Luttinger's theorem, where $\delta$ is the charge-carrier doping concentration. (b) Alternatively, based on the charge-spin separation, the analytical calculation to implement the elimination of the double electron occupancy is the slave-particle approach \cite{Feng15,Lee06,Yu92}, where the physical electron is decoupled according to its charge and spin degrees of freedom, and then the no-double electron occupancy constraint can be treated at {\it all temperatures}. However, a long-standing problem is how a microscopic SC theory based on the charge-spin separation can give a consistent description of a large EFS in cuprate superconductors. In our recent work \cite{Feng15a}, we follow the kinetic-energy-driven SC mechanism to develop a full charge-spin recombination scheme, where a charge carrier and a localized spin are fully recombined into a physical electron. Within this full charge-spin recombination scheme, we \cite{Feng15a} have studied the electronic structure of cuprate superconductors in the SC-state. For a superconductor, EFS is defined just above the SC transition temperature $T_{\rm c}$. However, we \cite{Feng15a} have defined operationally {\it EFS} of cuprate superconductors in the SC-state as the contour determined from the low-energy spectral weight, and then show that EFS of cuprate superconductors in the SC-state is a large EFS in the entire doping range, while the anomalous peak-dip-hump structure in the electron spectrum is mainly caused by the pseudogap, in qualitative agreement with the corresponding ARPES experimental results \cite{Hashimoto15}. In this paper, we try to discuss the microscopic origin of charge order and its evolution with doping in the pseudogap phase along with this line. As a complement of the our previous analysis of EFS of cuprate superconductors in the SC-state \cite{Feng15a}, we first study the nature of EFS in the normal-state, and then show that the area enclosed by EFS is identical to the total numbers of electrons as expected from Luttinger's theorem. However, the antinodal region of EFS is suppressed by the electron self-energy in the particle-hole channel, leading to that EFS is broken into the disconnected Fermi arcs around the nodal region. In particular, the charge-order state is driven by the Fermi-arc instability, with a characteristic wave vector corresponding to the hot spots of the Fermi arcs rather than the antinodal nesting vector. Our theory also shows that the Fermi arc, charge order, and pseudogap in cuprate superconductors are intimately related each other, and all of them emanates from the electron self-energy in the particle-hole channel due to the interaction between electrons by the exchange of spin excitations.

The rest of this paper is organized as follows. We present the basic formalism in Section \ref{Formalism}, where we reduce the electron Green's function of cuprate superconductors in the SC-state obtained from the full charge-spin recombination scheme to the present case in the normal-state. The quantitative characteristics of the nature of charge order and its evolution with doping in the pseudogap phase then is discussed in Section \ref{characteristics}, where we show that although the Fermi arc increases its length as a function of doping, the charge-order wave vector $Q_{\rm CO}$ reduces linearity with the increase of doping. Finally, we give a summary in Section \ref{conclusions}.

\section{Formalism}\label{Formalism}

Very soon after the discovery of superconductivity in cuprate superconductors \cite{Bednorz86}, Anderson \cite{Anderson87} suggested that the essential physics of cuprate superconductors can be described by the $t$-$J$ model on a square lattice,
\begin{eqnarray}\label{tJmodel}
H=-\sum_{lm\sigma}t_{lm}C^{\dagger}_{l\sigma}C_{m\sigma}+\mu\sum_{l\sigma}C^{\dagger}_{l\sigma}C_{l\sigma}+J\sum_{\langle ll'\rangle}{\bf S}_{l}\cdot {\bf S}_{l'},~~
\end{eqnarray}
where the summation is over all sites $l$, and the hopping amplitudes $t_{lm}$ connect sites $l$ and $m$. In the kinetic-energy term, we will restrict to our attention to the nearest ($t$) and next nearest ($-t'$) neighbor hoppings, while $\langle ll'\rangle$ in the Heisenberg term means the spin-spin interaction occurs only for the nearest-neighbor (NN) sites. $C^{\dagger}_{l\sigma}$ ($C_{l\sigma}$) denotes the electron creation (destruction) operator of one electron on site $l$ with spin $\sigma=\uparrow,\downarrow$, ${\bf S}_{l}=(S^{\rm x}_{l},S^{\rm y}_{l}, S^{\rm z}_{l})$ are spin operators, and $\mu$ is the chemical potential. In this $t$-$J$ model (\ref{tJmodel}), the hopping amplitudes $t_{lm}$ measure the electron delocalization through the lattice, while the AF exchange interaction $J$ describes AF coupling between localized spins. In spite of its simple form, the $t$-$J$ model (\ref{tJmodel}) has been proved to be very difficult to analyze, analytically as well as numerically, because of the restriction of the motion of electrons in the restricted Hilbert space without double electron occupancy, i.e., $\sum_{\sigma}C^{\dagger}_{l\sigma} C_{l\sigma}\leq 1$. In the early days of superconductivity, we \cite{Feng9404,Feng15} have developed a fermion-spin theory to confront this no-double electron occupancy constraint, where the constrained electron operators $C_{l\uparrow}$ and $C_{l\downarrow}$ are decoupled as,
\begin{eqnarray}\label{css}
C_{l\uparrow}=h^{\dagger}_{l\uparrow}S^{-}_{l},~~~~C_{l\downarrow}=h^{\dagger}_{l\downarrow}S^{+}_{l},
\end{eqnarray}
respectively, with the fermion operator $h_{l\sigma}=e^{-i\Phi_{l\sigma}}h_{l}$ that keeps track of the charge degree of freedom together with some effects of spin configuration rearrangements due to the presence of the doped hole itself (charge carrier), while the spin operator $S_{l}$ represents the spin degree of freedom, and then the local constraint of no double electron occupancy is always satisfied in actual calculations. In this fermion-spin representation (\ref{css}), the original $t$-$J$ model (\ref{tJmodel}) can be rewritten as,
\begin{eqnarray}\label{cssham}
H&=&\sum_{lm}t_{lm}(h^{\dagger}_{m\uparrow}h_{l\uparrow}S^{+}_{l}S^{-}_{m}+h^{\dagger}_{m\downarrow}h_{l\downarrow}S^{-}_{l}S^{+}_{m})\nonumber\\
&-&\mu\sum_{l\sigma}h^{\dagger}_{l\sigma} h_{l\sigma} +J_{{\rm eff}}\sum_{\langle ll'\rangle}{\bf S}_{l}\cdot {\bf S}_{l'},~~~~
\end{eqnarray}
where $S^{-}_{l}=S^{\rm x}_{l}-iS^{\rm y}_{l}$ and $S^{+}_{l}=S^{\rm x}_{l}+iS^{\rm y}_{l}$ are the spin-lowering and spin-raising operators for the spin $S=1/2$, respectively, $J_{{\rm eff}}=(1-\delta)^{2}J$, and $\delta=\langle h^{\dagger}_{l\sigma}h_{l\sigma}\rangle=\langle h^{\dagger}_{l}h_{l}\rangle$ is the charge-carrier doping concentration. In conventional metals, the fermiology is essentially determined by the lowering of the total kinetic energy which becomes possible when a periodic potential supports the delocalization of the electron local state into extended one with a well-defined momentum and a correspondingly homogeneous distribution of the charge density \cite{Comin15}. However, in doped cuprates, the restriction of no double electron occupancy in a given site induces a strong coupling between the charge and spin degrees of freedom of the constrained electron, which reflects a fact that even the kinetic energy in the $t$-$J$ model (\ref{tJmodel}) has strong Coulombic contribution, and therefore pushes the system to find new ways to lower its total energy. This tendency on the other hand leads to the emergence of a rich variety of the doping dependence of ordered states \cite{Comin15}.

Based on the $t$-$J$ model (\ref{cssham}) in the fermion-spin representation, the kinetic-energy-driven SC mechanism has been developed in the doped regime without an AF long-range order \cite{Feng15,Feng0306,Feng12}, where the interaction between charge carriers directly from the kinetic energy of the $t$-$J$ model (\ref{cssham}) by the exchange of spin excitations induces the d-wave SC-state in the particle-particle channel and pseudogap state in the particle-hole channel, therefore there is a coexistence of the SC-state and
pseudogap state in the whole SC dome. The electron Cooper pairs on the other hand originated from the charge-carrier pairing state are due to the charge-spin recombination. For the discussions of the electronic structure of cuprate superconductors in the SC-state, we \cite{Feng15a} have developed a full charge-spin recombination scheme, where the coupling form between the electron quasiparticle and spin excitation in the $t$-$J$ model (\ref{cssham}) is the same as that between the charge-carrier quasiparticle and spin excitation, and then the full electron diagonal and off-diagonal Green's functions of cuprate superconductors in the SC-state are obtained. Following our previous discussions \cite{Feng15a}, the full electron Green's function of cuprate superconductors in the SC-state obtained from the full charge-spin recombination scheme can be reduced to the present case in the normal-state by the condition of the SC gap parameter $\bar{\Delta}=0$,
\begin{eqnarray}\label{EGF}
G({\bf k},\omega)={1\over \omega-\varepsilon_{\bf k}-\Sigma_{1}({\bf k},\omega)},
\end{eqnarray}
where the mean-field (MF) electron Green's function $G^{(0)-1}({\bf k},\omega)=\omega-\varepsilon_{\bf k}$, the MF electron excitation spectrum $\varepsilon_{\bf k}=-Zt\gamma_{\bf k} +Zt'\gamma_{\bf k}'+\mu$, with $\gamma_{\bf k}=({\rm cos}k_{x}+{\rm cos}k_{y})/2$, $\gamma_{\bf k}'= {\rm cos} k_{x}{\rm cos}k_{y}$, and $Z$ is the number of the NN or next NN sites on a square lattice, while the electron self-energy $\Sigma_{1}({\bf k},\omega)$ in the particle-hole channel is obtained directly from the corresponding electron self-energy in the SC-state given in Ref. \cite{Feng15a} in the case of $\bar{\Delta}=0$,
\begin{eqnarray}\label{ESE}
\Sigma_{1}({\bf k},i\omega_{n})&=&{1\over N^{2}}\sum_{{\bf p,p'}}\Lambda^{2}_{{\bf p}+{\bf p}'+{\bf k}}\nonumber\\
&\times& {1\over \beta}\sum_{ip_{m}}G({\bf p}+{\bf k},ip_{m}+i\omega_{n})\Pi({\bf p},{\bf p}',ip_{m}),~~~~~
\end{eqnarray}
with $\Lambda_{{\bf k}}=Zt\gamma_{\bf k}-Zt'\gamma_{\bf k}'$, and the spin bubble,
\begin{eqnarray}\label{SB}
\Pi({\bf p},{\bf p}',ip_{m})&=&{1\over\beta}\sum_{ip'_{m}}D^{(0)}({\bf p'},ip_{m}')\nonumber\\
&\times&D^{(0)}({\bf p}'+{\bf p},ip_{m}'+ip_{m}),
\end{eqnarray}
where the MF spin Green's function has been given by,
\begin{eqnarray}\label{MFSGF}
D^{(0)}({\bf k},\omega)={B_{\bf k}\over 2\omega_{\bf k}}\left ({1\over \omega-\omega_{\bf k}}-{1\over\omega+\omega_{\bf k}}\right ),
\end{eqnarray}
with the MF spin excitation spectrum $\omega_{\bf k}$ and function $B_{\bf k}$ that have been given explicitly in Ref. \cite{Feng15}. In cuprate superconductors, the strong electron correlation and the related quasiparticle coherence are closely related to the electron self-energy $\Sigma_{1}({\bf k},\omega)$, and then the behavior of the electrons in the normal-state is most fully described in terms of the electron Green's function, the poles of which map the energy versus momentum dependence of the electron quasiparticles. The locus of the poles at the electron Fermi energy defines EFS, from which almost all the electronic properties in the normal-state emanate. However, in the framework of the full charge-spin recombination \cite{Feng15a}, this electron self-energy $\Sigma_{1}({\bf k},\omega)$ in Eq. (\ref{ESE}) originates in the electron's coupling to spin excitations, and therefore in striking contrast to the electron self-energy in conventional superconductors, $\Sigma_{1}({\bf k},\omega)$ is strong dependence of energy and momentum \cite{Feng15a}, which therefore complicates the physical properties of the electronic structure in cuprate superconductors.

Eqs. (\ref{EGF}) and (\ref{ESE}) show that the electron self-energy $\Sigma_{1}({\bf k},\omega)$ and electron Green's function $G({\bf k},\omega)$ are related self-consistently. However, the electron self-energy $\Sigma_{1}({\bf k},\omega)$ in Eq. (\ref{ESE}) is not an even function of $\omega$. For the evaluation of $\Sigma_{1}({\bf k},\omega)$, we break up $\Sigma_{1}({\bf k},\omega)$ as $\Sigma_{1}({\bf k},\omega)=\Sigma_{\rm 1e} ({\bf k},\omega)+\omega\Sigma_{\rm 1o}({\bf k},\omega)$, with $\Sigma_{\rm 1e}({\bf k},\omega)$ and $\omega\Sigma_{\rm 1o}({\bf k},\omega) $ that are the corresponding symmetric and antisymmetric parts of $\Sigma_{1}({\bf k},\omega)$, respectively, and then both $\Sigma_{\rm 1e}({\bf k}, \omega)$ and $\Sigma_{\rm 1o} ({\bf k},\omega)$ are even functions of $\omega$. In particular, the antisymmetric part $\Sigma_{\rm 1o}({\bf k},\omega)$ of the electron self-energy $\Sigma_{1}({\bf k},\omega)$ is directly related to the electron quasiparticle coherent weight as \cite{Mahan81} $Z^{-1}_{\rm F}({\bf k},\omega)=1-{\rm Re}\Sigma_{\rm 1o}({\bf k},\omega)$. As a first step of discussions, we only focus on the low-energy behavior. In this case, the electron quasiparticle coherent weight can be generally discussed in the static limit, i.e., $Z^{-1}_{\rm F} ({\bf k})=1-{\rm Re} \Sigma_{\rm 1o}({\bf k},\omega)\mid_{\omega=0}$. As in conventional superconductors \cite{Mahan81}, the retarded function ${\rm Re}\Sigma_{\rm 1e}({\bf k}, \omega) \mid_{\omega=0}$ just renormalizes the chemical potential. Although $Z_{\rm F}({\bf k})$ still is a function of momentum, the momentum dependence may be unimportant in a qualitative discussion, and therefore the wave vector ${\bf k}$ in $Z_{\rm F}({\bf k})$ can be chosen as,
\begin{eqnarray}\label{ESCE1}
{1\over Z_{\rm F}}=1-{\rm Re}\Sigma_{\rm 1o}({\bf k}_{\rm A},\omega=0)\mid_{{\bf k}=[\pi,0]},
\end{eqnarray}
just as it has been done in the ARPES experiments \cite{Ding01,DLFeng00}. Moreover, this electron quasiparticle coherent weight $Z_{\rm F}$ reduces the electron quasiparticle bandwidth, and suppresses the spectral weight of the single-particle excitation spectrum \cite{Feng15a}. With the above static-limit approximation, the full electron Green's function in Eq. (\ref{EGF}) can be evaluated explicitly as,
\begin{eqnarray}\label{MFEGF}
G({\bf k},\omega)=Z_{\rm F}{1\over\omega-\bar{\varepsilon}_{\bf k}},
\end{eqnarray}
with $\bar{\varepsilon}_{\bf k}=Z_{\rm F}\varepsilon_{\bf k}$. Although the form of the electron Green's function in Eq. (\ref{MFEGF}) obtained in the static-limit approximation is similar to that obtained in the MF approximation, the partial effect of the quasiparticle coherence has been contained in terms of the quasiparticle coherent weight $Z_{\rm F}$.

With the help of the electron Green's function in Eq. (\ref{MFEGF}) and spin Green's function in Eq. (\ref{MFSGF}), the electron self-energy $\Sigma_{1}({\bf k}, \omega)$ in Eq. (\ref{ESE}) is evaluated explicitly as,
\begin{eqnarray}\label{ESE1}
\Sigma_{1}({\bf k},\omega)&=&{1\over N^{2}}\sum_{{\bf pp'}\mu\nu}(-1)^{\nu+1}\Omega_{\bf pp'k}\nonumber\\
&\times&{F^{(\nu)}_{{\rm n\mu} {\bf pp'k}}\over\omega+(-1)^{\mu+1}\omega_{\nu{\bf p}{\bf p}'}- \bar{\varepsilon}_{{\bf p}+{\bf k}}},~~~
\end{eqnarray}
with $\mu (\nu)=1,2$, $\omega_{\nu{\bf p}{\bf p}'}=\omega_{{\bf p}+{\bf p}'}-(-1)^{\nu}\omega_{\bf p'}$, $\Omega_{\bf pp'k}=Z_{\rm F}\Lambda^{2}_{{\bf p}+{\bf p}'+{\bf k}}B_{{\bf p}'} B_{{\bf p}+{\bf p}'}/(4\omega_{{\bf p}'}\omega_{{\bf p}+{\bf p}'})$, and the function,
\begin{eqnarray}
F^{(\nu)}_{{\rm n\mu}{\bf pp'k}}=n_{\rm F}[(-1)^{\mu+1}\bar{\varepsilon}_{{\bf p}+{\bf k}}]n^{(\nu)}_{{\rm 1B} {\bf pp'}}+n^{(\nu)}_{{\rm 2B}{\bf pp'}},
\end{eqnarray}
where $n^{(\nu)}_{{\rm 1B}{\bf pp'}}=1+n_{\rm B}(\omega_{{\bf p}'+{\bf p}})+n_{\rm B}[(-1)^{\nu+1}\omega_{\bf p'}]$, $n^{(\nu)}_{{\rm 2B}{\bf pp'}}=n_{\rm B}(\omega_{{\bf p}'+{\bf p}}) n_{\rm B}[(-1)^{\nu+1} \omega_{\bf p'}]$, and $n_{\rm B}(\omega)$ and $n_{\rm F}(\omega)$ are the boson and fermion distribution functions, respectively. In this case, the electron quasiparticle coherent weight $Z_{\rm F}$ satisfies the following self-consistent equation,
\begin{eqnarray}\label{ESE3}
{1\over Z_{\rm F}}&=&1+{1\over N^{2}}\sum_{{\bf pp'}\nu}(-1)^{\nu+1}\Omega_{{\bf pp'}{\bf k}_{\rm A}}\left [{F^{(\nu)}_{{\rm n1}{\bf pp'}{\bf k}_{\rm A}}\over (\omega_{\nu{\bf p}{\bf p}'}- \bar{\varepsilon}_{{\bf p}+{\bf k}_{\rm A}})^{2}}\right . \nonumber\\
&+&\left . {F^{(\nu)}_{{\rm n2}{\bf pp'}{\bf k}_{\rm A}}\over (\omega_{\nu{\bf p}{\bf p}'}+ \bar{\varepsilon}_{{\bf p}+{\bf k}_{\rm A}})^{2}} \right ],~~~
\end{eqnarray}
where the wave vector ${\bf k}_{\rm A}=[\pi,0]$. The above equation (\ref{ESE3}) must be solved simultaneously with the following self-consistent equation,
\begin{eqnarray}\label{ESCE2}
1-\delta ={1\over N}\sum_{{\bf k}}Z_{\rm F}\left (1-{\rm tanh}[{1\over 2}\beta\bar{\varepsilon}_{\bf k}] \right ),
\end{eqnarray}
and then the electron quasiparticle coherent weight $Z_{\rm F}$ and chemical potential $\mu$ are determined by the self-consistent calculation without using any adjustable parameters.

\section{Charge order and its evolution with doping}\label{characteristics}

In this section, we show that the quantitative characteristics of charge order and its evolution with doping in the pseudogap phase observed from ARPES experiments \cite{Comin15,Comin14} can be described within the framework of the fermion-spin theory. In cuprate superconductors, although the values of $J$, $t$, and $t'$ are believed to vary somewhat from compound to compound \cite{Tanaka04}, however, as in our previous studies \cite{Feng15a}, the commonly used parameters in this paper are chosen as $t/J=2.5$ and $t'/t=0.3$ for a qualitative discussion.

\subsection{Electron Fermi surface}\label{Fermi-surface}

\begin{figure}[h!]
\centering
\includegraphics[scale=0.6]{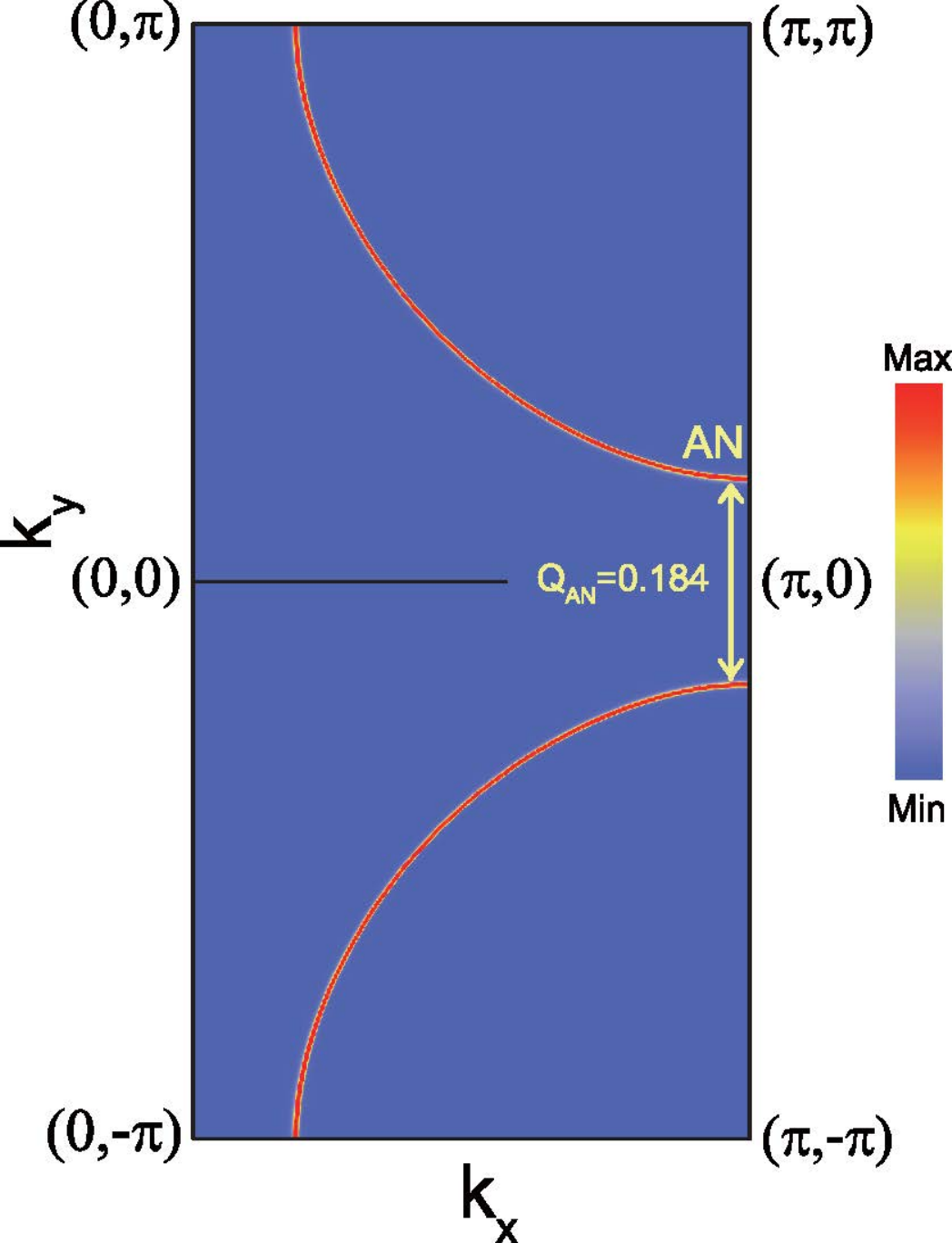}
\caption{(Color online) The map of the electron spectral intensity $A({\bf k},0)$ at $\delta=0.12$ with $T=0.002J$ for $t/J=2.5$ and $t'/t=0.3$. \label{spectral-maps}}
\end{figure}

\begin{figure}[h!]
\centering
\includegraphics[scale=0.6]{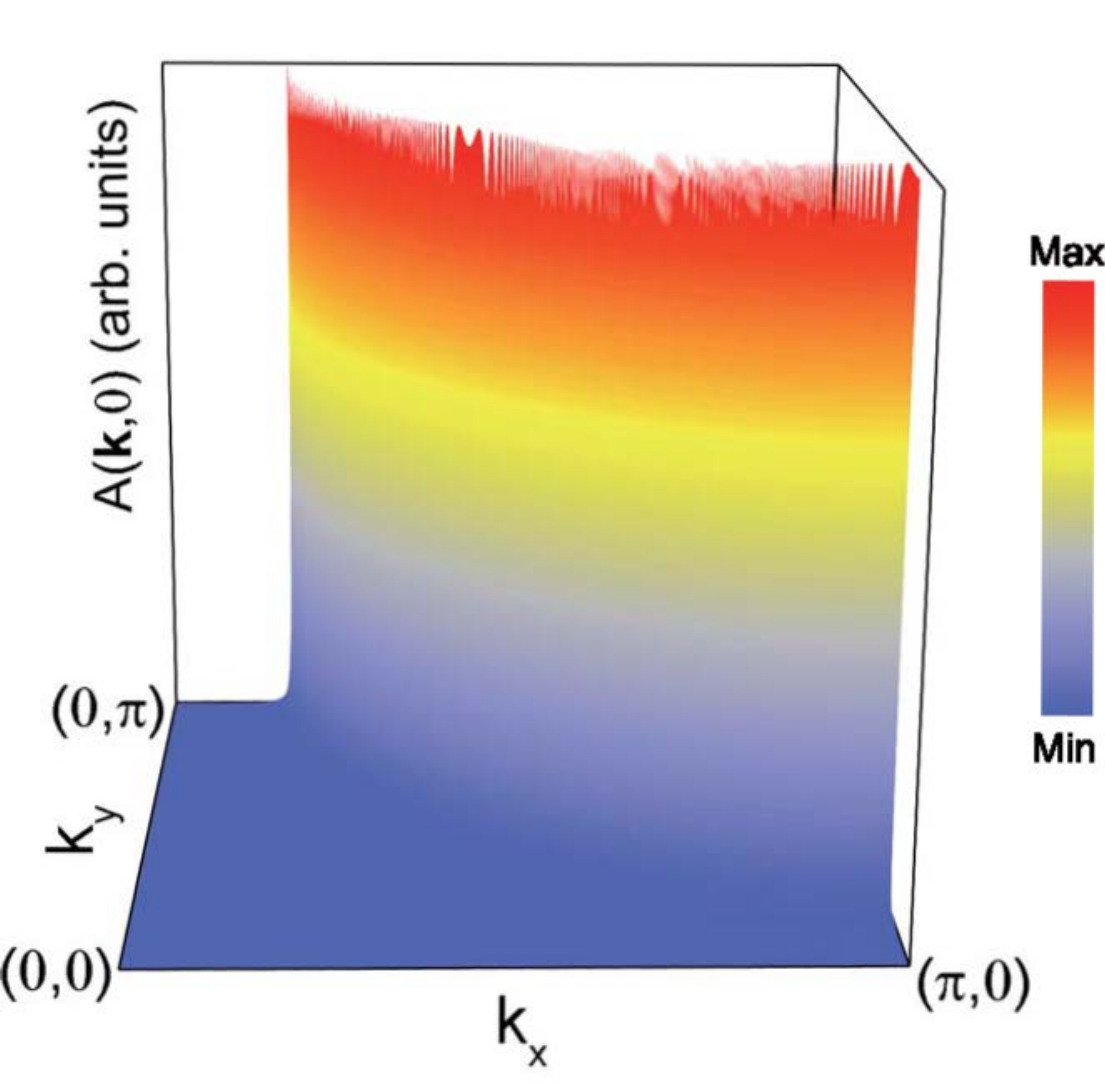}
\caption{(Color online) The electron spectral function $A({\bf k},0)$ in the $[k_{x},k_{y}]$ plane at $\delta=0.12$ with $T=0.002J$ for $t/J=2.5$ and $t'/t=0.3$. \label{spectral-maps-3D}}
\end{figure}

In ARPES experiments, the intensity of ARPES spectra at zero energy is used to map out the underlying EFS, i.e., the underlying EFS is determined by looking at the electron spectral function $A({\bf k},\omega=0)$ to map out the locus of the maximum in the intensity of $A({\bf k},\omega=0)$. The notion of EFS is one of the characteristic concepts in the field of condensed matter physics, and it plays a crucial role in the understanding of the physical properties of interacting electron systems. This is why a central question in the theory of cuprate superconductors concerns the nature and topology of EFS. For a better understanding of the intrinsic link between charge order and fermiology, we first discuss EFS in the static-limit approximation for the electron self-energy $\Sigma_{1}({\bf k}, \omega)$. In this case, the electron spectral function $A({\bf k},\omega)=-2{\rm Im}G({\bf k},\omega)$ is obtained directly from the above electron Green's function (\ref{MFEGF}) as,
\begin{eqnarray}\label{spectrum}
A({\bf k},\omega)=2\pi Z_{\rm F}\delta(\omega-\bar{\varepsilon}_{\bf k}),
\end{eqnarray}
where the energy and momentum dependence in the electron self-energy $\Sigma_{1}({\bf k},\omega)$ has been dropped, and then the electron spectral function in Eq. (\ref{spectrum}) has a Dirac delta function form on the quasiparticle dispersion curves, $\bar{\varepsilon}_{\bf k}$ versus ${\bf k}$, reflecting that the quasiparticles are {\it free} electrons. In other words, the form of the electron spectral function in Eq. (\ref{spectrum}) is similar to that obtained in the MF approximation, and then EFS is necessarily a surface in momentum-space on which the electron lifetime becomes infinitely long in the limit as one approaches EFS. In Fig. \ref{spectral-maps}, we plot a map of the electron spectral intensity $A({\bf k},0)$ in Eq. (\ref{spectrum}) for the underdoping $\delta=0.12$ with temperature $T=0.002J$, where EFS forms a continuous contour in momentum space. To show this point clearly, we plot $A({\bf k},0)$ in the $[k_{x},k_{y}]$ plane at $\delta=0.12$ with $T=0.002J$ in Fig. \ref{spectral-maps-3D}, where we see that the peaks with the same height distribute uniformly along EFS, and then a large EFS is formed as a closed contour of the gapless excitations in momentum space, in qualitative agreement with the early ARPES experimental results \cite{Takahashi89,Campuzano90,Olson90,Marshall96,Loeser96,Ding97}. Moreover, according to one of the self-consistent equations (\ref{ESCE2}), EFS with the area contains $1-\delta$ electrons, and therefore is consistent with that predicted by the Luttinger's theorem. However, the antinodal nesting, marked by the yellow arrow in Fig. \ref{spectral-maps}, yields an ordered wave vector $Q_{\rm AN}\sim 0.184$, which is in disagreement with the experimental average values \cite{Comin15,Wu11,Chang12,Ghiringhelli12,Ghiringhelli14,Comin14,Neto14,Hashimoto15,Comin15a} of the charge-order wave vector $Q_{\rm CD}\sim 0.256$ observed on cuprate superconductors at the doping concentration $\delta\approx 0.12$, reflecting a fact that the electron-hole scattering between antinodal excitations with the wave vector $Q_{\rm AN}$ is inadequate for a description of the charge-order state \cite{Comin14}. However, we will show in the following discussions that in the underdoped and optimally doped regimes, the energy and momentum dependence of the electron self-energy $\Sigma_{1}({\bf k},\omega)$ truncates this continuous contour in momentum space into the disconnected Fermi arcs located around the nodal region of the Brillouin zone, then the charge-order state is driven by the Fermi-arc instability, with a characteristic wave vector $Q_{\rm HS}$ corresponding to the hot spots of the Fermi arcs that is in agreement with the experimental average values of the charge-order wave vector $Q_{\rm CD}$ \cite{Comin15,Wu11,Chang12,Ghiringhelli12,Ghiringhelli14,Comin14,Neto14,Hashimoto15,Comin15a}.

\subsection{Charge order driven by Fermi-arc instability}\label{self-energy-Fermi-surface}

\begin{figure}[h!]
\centering
\includegraphics[scale=0.6]{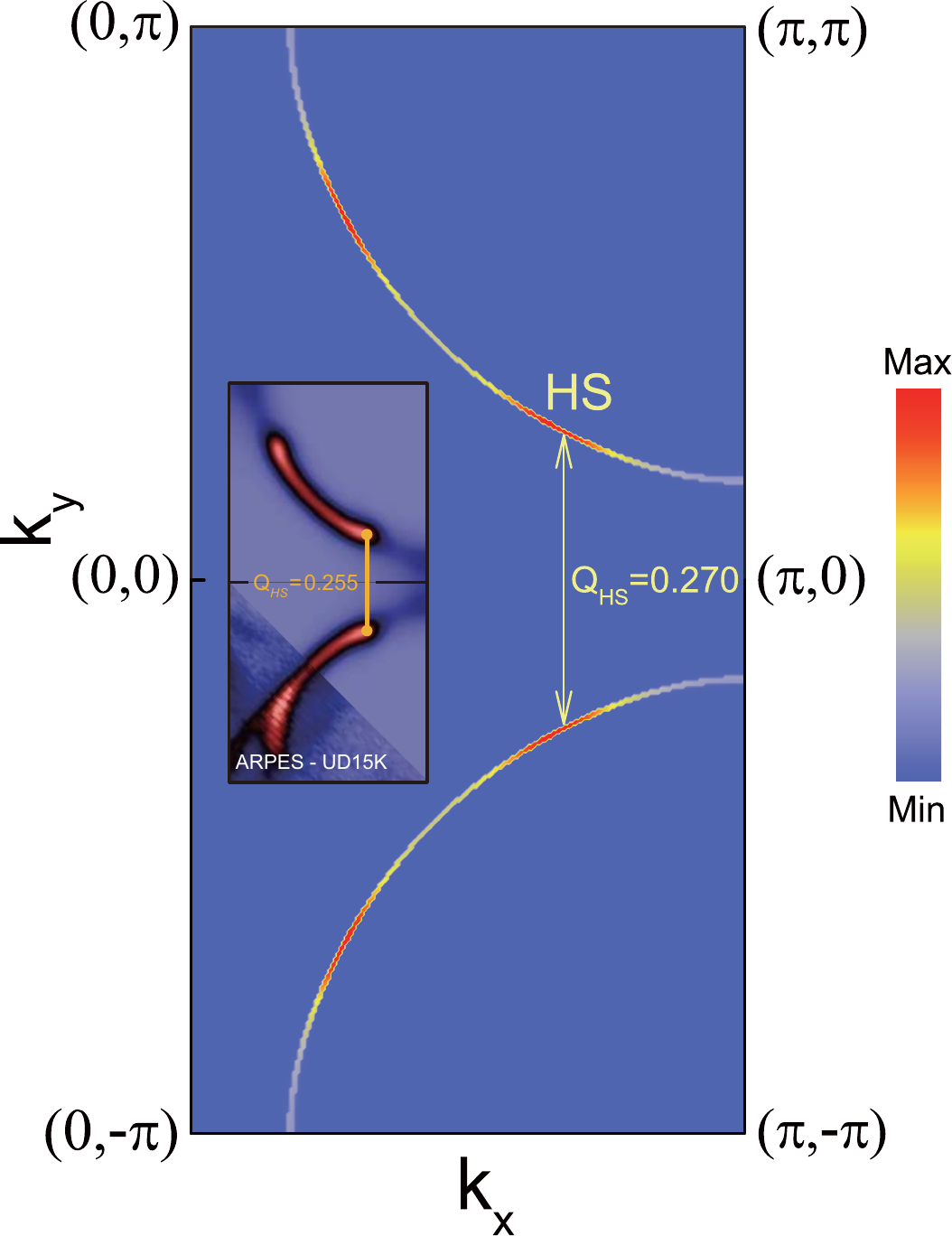}
\caption{(Color online) The map of the electron spectral intensity $A({\bf k},0)$ at $\delta=0.12$ with $T=0.002J$ for $t/J=2.5$ and $t'/t=0.3$. Inset: the corresponding experimental data of the underdoped Bi$_{2}$Sr$_{2-x}$La$_{x}$CuO$_{6+\delta}$ taken from Ref. \cite{Comin14}. \label{SE-spectral-maps}}
\end{figure}

\begin{figure}[h!]
\centering
\includegraphics[scale=0.6]{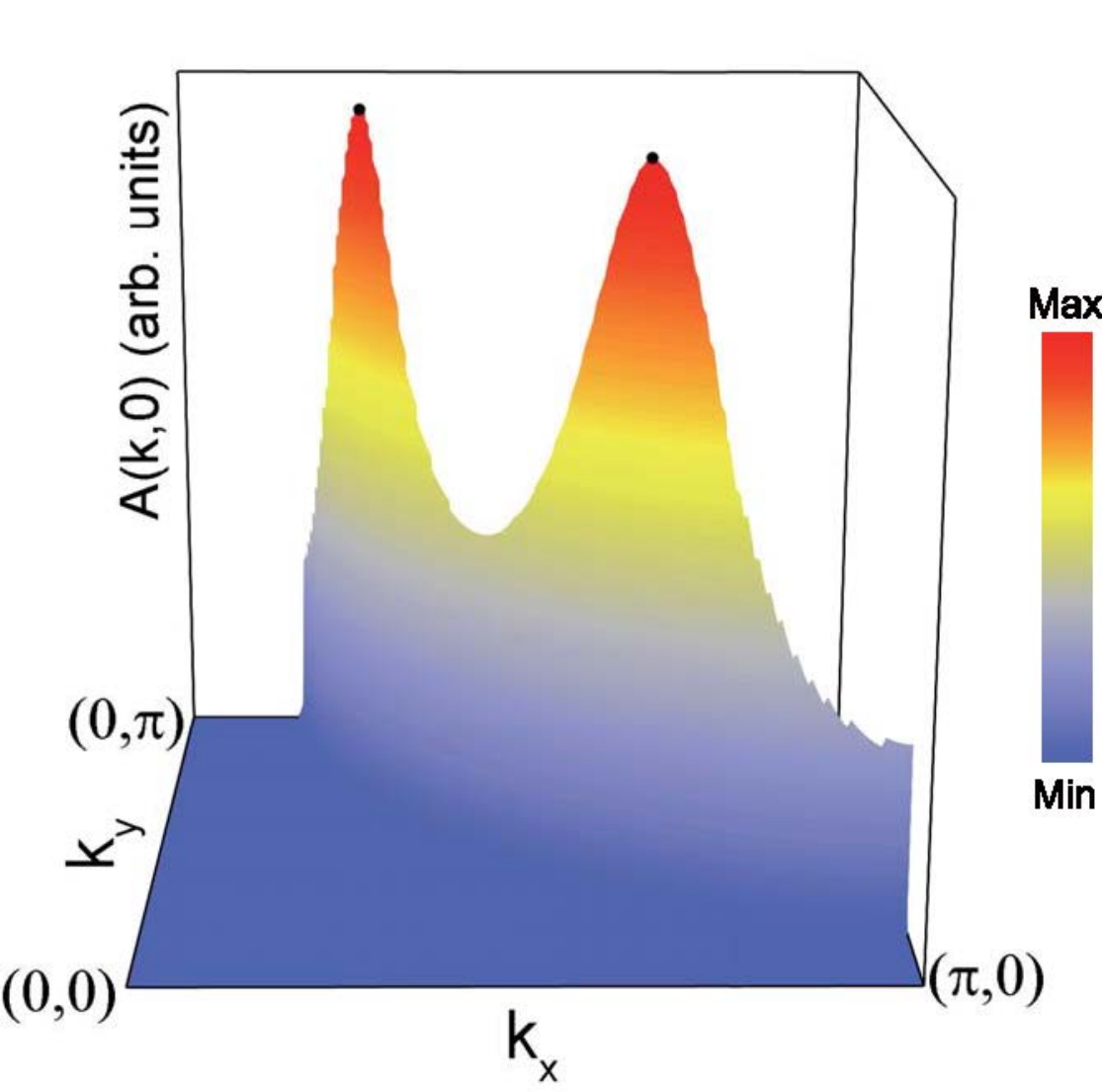}
\caption{(Color online) The electron spectral function $A({\bf k},0)$ in the $[k_{x},k_{y}]$ plane at $\delta=0.12$ with $T=0.002J$ for $t/J=2.5$ and $t'/t=0.3$. The black circles indicate the locations of the hot spots. \label{SE-spectral-maps-3D}}
\end{figure}

As a doped Mott insulator, the strong electron correlation in cuprate superconductors is closely related to the electron self-energy $\Sigma_{1}({\bf k},\omega)$. In particular, the detailed form of the line shape of the electron spectrum and reconstruction of EFS are determined by $\Sigma_{1}({\bf k},\omega)$. In order to take into account the effect of $\Sigma_{1}({\bf k},\omega)$ on the electronic structure, the electron spectral function can be evaluated directly from the electron Green's function (\ref{EGF}) as,
\begin{eqnarray}\label{full-spectral-function}
A({\bf k},\omega)={2|{\rm Im}\Sigma_{1}({\bf k},\omega)|\over [\omega-\varepsilon_{\bf k}-{\rm Re}\Sigma_{1}({\bf k},\omega)]^{2}+[{\rm Im}\Sigma_{1}({\bf k},\omega)]^{2}},~~~
\end{eqnarray}
where ${\rm Im}\Sigma_{1}({\bf k},\omega)$ and ${\rm Re}\Sigma_{1}({\bf k},\omega)$ are the corresponding imaginary and real parts of $\Sigma_{1}({\bf k},\omega)$ in Eq. (\ref{ESE1}), respectively. Since the electron self-energy $\Sigma_{1}({\bf k},\omega)$ is involved in the electron spectral function (\ref{full-spectral-function}), the electron quasiparticle energies are renormalized and they acquire a finite lifetime. In Fig. \ref{SE-spectral-maps}, we plot a map of the electron spectral intensity $A({\bf k},0)$ in Eq. (\ref{full-spectral-function}) at $\delta=0.12$ with $T=0.002J$ in comparison with the corresponding experimental result \cite{Comin14} obtained from  Bi$_{2}$Sr$_{2-x}$La$_{x}$CuO$_{6+\delta}$ at the doping regime around $\delta\approx 0.12$ (inset). Comparing it with Fig. \ref{spectral-maps} for the same set of parameters except for the effect from the energy and momentum dependence of $\Sigma_{1}({\bf k},\omega)$, we see that EFS around the antinodal region has been suppressed by the electron self-energy $\Sigma_{1}({\bf k},\omega)$, leaving behind disconnected Fermi arcs centered around the nodal region, in qualitative agreement with the experimental results \cite{Comin14,Norman98,Kanigel06,Yoshida06,Tanaka06,Kanigel07,Nakayama09,Yoshida09,Meng11,Ideta12,Kondo13}. For a determination of the positions of the hot spots on the Fermi arcs, we plot $A({\bf k},0)$ in the $[k_{x},k_{y}]$ plane at $\delta=0.12$ with $T=0.002J$ in Fig. \ref{SE-spectral-maps-3D}, where the locations of the hot spots, marked by the black circles, are thus determined by the highest peak heights on the Fermi arcs. In this case, the wave vector connecting the hot spots is found to be $Q_{\rm HS}\sim 0.270$, closely matching the experimental average values of the charge-order wave vector $Q_{\rm CD}\sim 0.256$ found in the underdoped cuprate superconductors \cite{Comin15,Wu11,Chang12,Ghiringhelli12,Ghiringhelli14,Comin14,Neto14,Hashimoto15,Comin15a}. These hot spots connected by the charge-order wave vector contribute effectively to the electron quasiparticle scattering process.

\subsection{Electron momentum distribution}\label{Electron-momentum-distribution}

\begin{figure}[h!]
\centering
\includegraphics[scale=0.65]{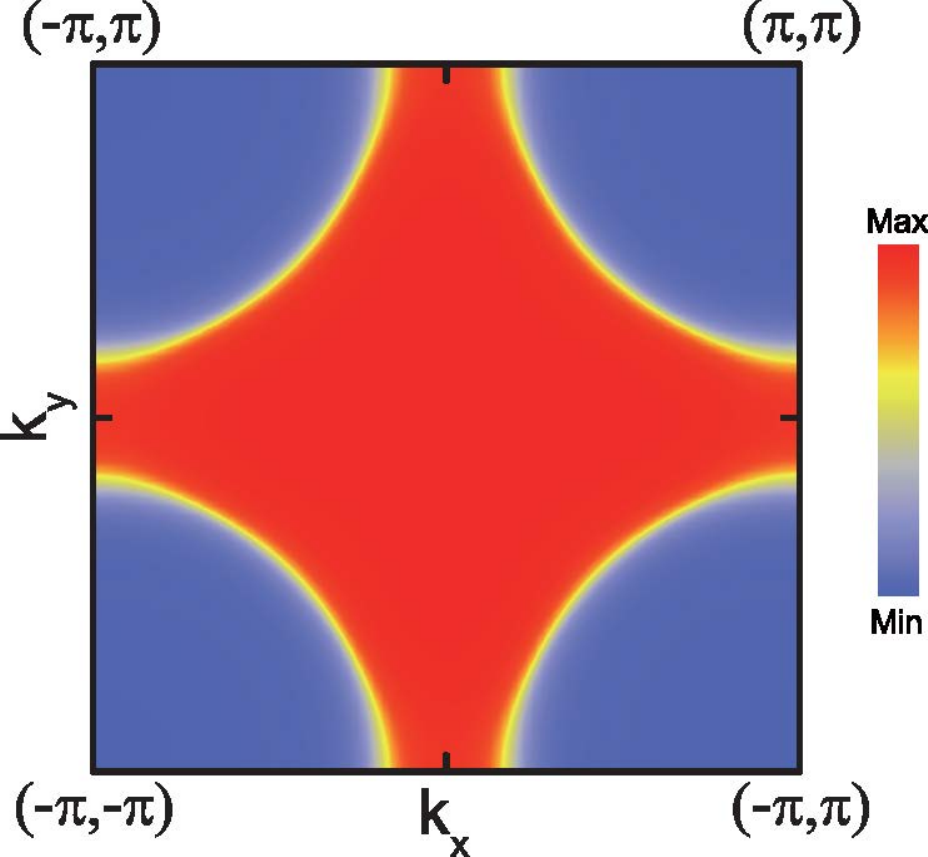}
\caption{(Color online) The map of the electron-momentum distribution over the entire Brillouin zone at $\delta=0.12$ with $T=0.002J$ for $t/J=2.5$ and $t'/t=0.3$. \label{map-SE-momentum-distribution}}
\end{figure}

\begin{figure}[h!]
\centering
\includegraphics[scale=0.45]{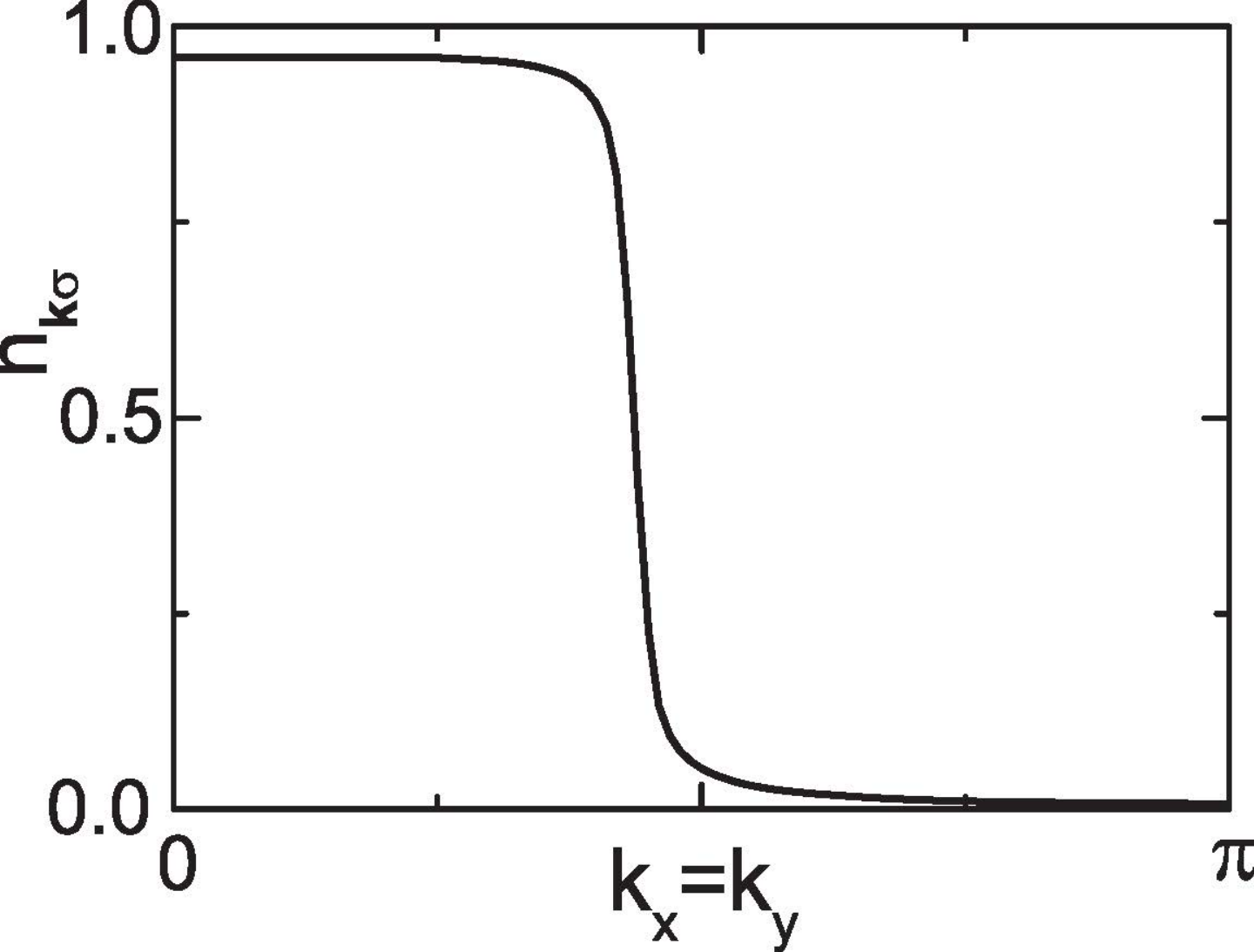}
\caption{The electron-momentum distribution along the ${\bf k}=[0,0]$ to ${\bf k}=[\pi,\pi]$ direction at $\delta=0.12$ with $T=0.002J$ for $t/J=2.5$ and $t'/t=0.3$. \label{SE-momentum-distribution}}
\end{figure}

The number of electrons in state {\bf k} is obtained by summing over all energies $\omega$, weight by the electron spectral function (\ref{full-spectral-function}) as,
\begin{eqnarray}\label{number}
n_{{\bf k}\sigma}= \int^{\infty}_{-\infty}{{\rm d}\omega\over 2\pi}n_{\rm F}(\omega)A({\bf k},\omega),
\end{eqnarray}
where $n_{{\bf k}\sigma}$ is so-called the electron momentum distribution. This electron momentum distribution is a quantity of great interest and its determination is very important, since EFS is also given by the set of $k$ values for which $n_{{\bf k}\sigma}$ shows a jump in discontinuity. When this discontinuity is smeared out, the gradient of $n_{{\bf k}\sigma}$, $\nabla n_{{\bf k}\sigma}$, is assumed to be maximal at the locus of the underlying EFS \cite{Gros06}. In Fig. \ref{map-SE-momentum-distribution}, we plot a map of $n_{{\bf k}\sigma}$ over the entire Brillouin zone at $\delta=0.12$ with $T=0.002J$, where the identification of a large EFS is unambiguous. To further analyze the nature of EFS, we plot $n_{{\bf k}\sigma}$ along the ${\bf k}=[0,0]$ to ${\bf k}=[\pi,\pi]$ direction at $\delta=0.12$ with $T=0.002J$ in Fig. \ref{SE-momentum-distribution}. For a non-interacting electron system, the electron states are filled up to the Fermi momentum $k_{\rm F}$, and then $n_{{\bf k}\sigma}$ shows a sudden drop at $k_{\rm F}$. However, when the electron interaction (then the electron correlation) increases, $n_{{\bf k} \sigma}$ begins to deform, and then the electrons that used to occupy states below $k_{\rm F}$ in the case without electron interaction have moved to the states that were unoccupied in the presence of the electron interaction. The present result in Fig. \ref{SE-momentum-distribution} thus shows that the shape of the electron momentum distribution of cuprate superconductors in the normal-state is a should-be electron momentum distribution, i.e., in some part (below the electron Fermi energy) the distribution is closer to $1$, while in other part (above the electron Fermi energy) it is approximately closer to zero, in qualitative agreement with the results obtained from the numerical simulations \cite{Stephan91,Dagotto92,Moreo90,Gros06,Paramekanti01,Edegger06} and experimental observation \cite{Randeria95,Campuzano04}. In particular, although the low-energy spectral weight has been renormalized by the electron self-energy $\Sigma_{1}({\bf k},\omega)$, the area contained by EFS is invariant under the interaction effect, and then the underlying EFS still fulfills Luttinger's theorem.

\subsection{Doping dependence of charge-order wave vector}\label{charge-order-wave-vector}

\begin{figure}[h!]
\centering
\includegraphics[scale=0.4]{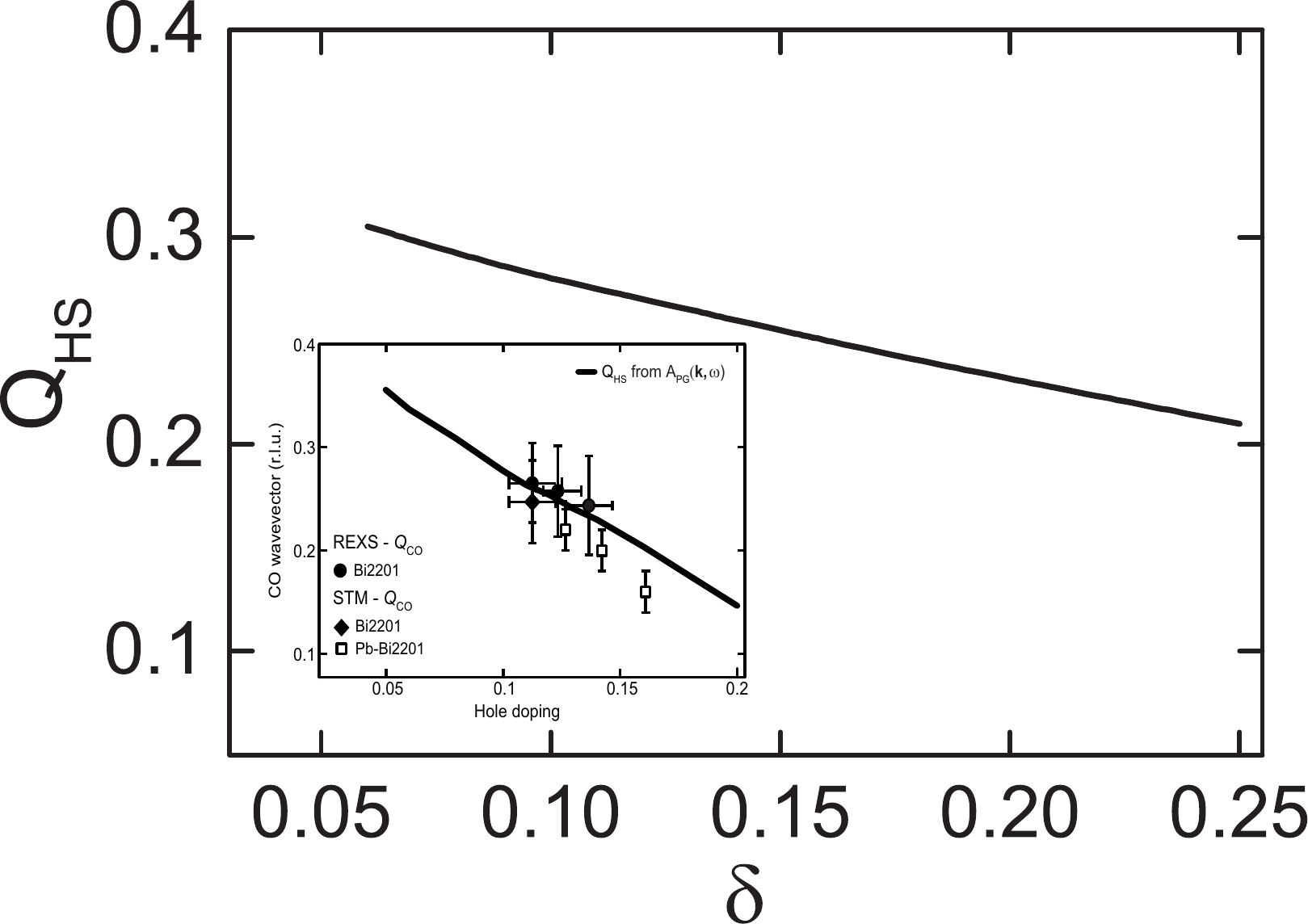}
\caption{The charge-order wave vector as a function of doping for $t/J=2.5$ and $t'/t=0.3$ with $T=0.002J$. Inset: the corresponding experimental data of Bi$_{2}$Sr$_{2-x}$La$_{x}$CuO$_{6+\delta}$ taken from Ref. \cite{Comin14}. \label{charge-order-wave-vector-doping}}
\end{figure}

Now we turn to discuss the evolution of charge order with doping. We have made a series of calculations for the electron spectral function $A({\bf k},0)$ at different doping concentrations, and the results show that although the Fermi-arc length is rather short in the underdoped regime, the Fermi arc increases its length as a function of doping, and therefore there is a tendency towards to form a closed contour in momentum space in the heavily overdoped regime \cite{Norman98,Kanigel06,Nakayama09,Yoshida06,Meng11,Ideta12}. However, in contrast to the case of the Fermi-arc length, the wave vector $Q_{\rm HS}$ smoothly decreases with the increase of doping. To show this evolution of $Q_{\rm HS}$ with doping clearly, the result for the extracted wave vector $Q_{\rm HS}$ as a function of doping with $T=0.002J$ is plotted in Fig. \ref{charge-order-wave-vector-doping} in comparison with the Bi$_{2}$Sr$_{2-x}$La$_{x}$CuO$_{6+\delta}$ experimental data \cite{Comin14} of the charge-order wave vector $Q_{\rm CO}$, where $Q_{\rm HS}$ reduces almost linearity with increasing doping. Incorporating both the results in Fig. \ref{SE-spectral-maps} and Fig. \ref{charge-order-wave-vector-doping}, it is thus shown that both the wave vector magnitude $Q_{\rm HS}$ and its evolution with doping qualitatively agree with the corresponding Bi$_{2}$Sr$_{2-x}$La$_{x}$CuO$_{6+\delta}$ experimental data \cite{Comin14}. The qualitative agreement between the present theoretical results and experimental data therefore suggests that the charge-order state is indeed driven by the Fermi-arc instability.

\begin{figure}[h!]
\centering
\includegraphics[scale=0.65]{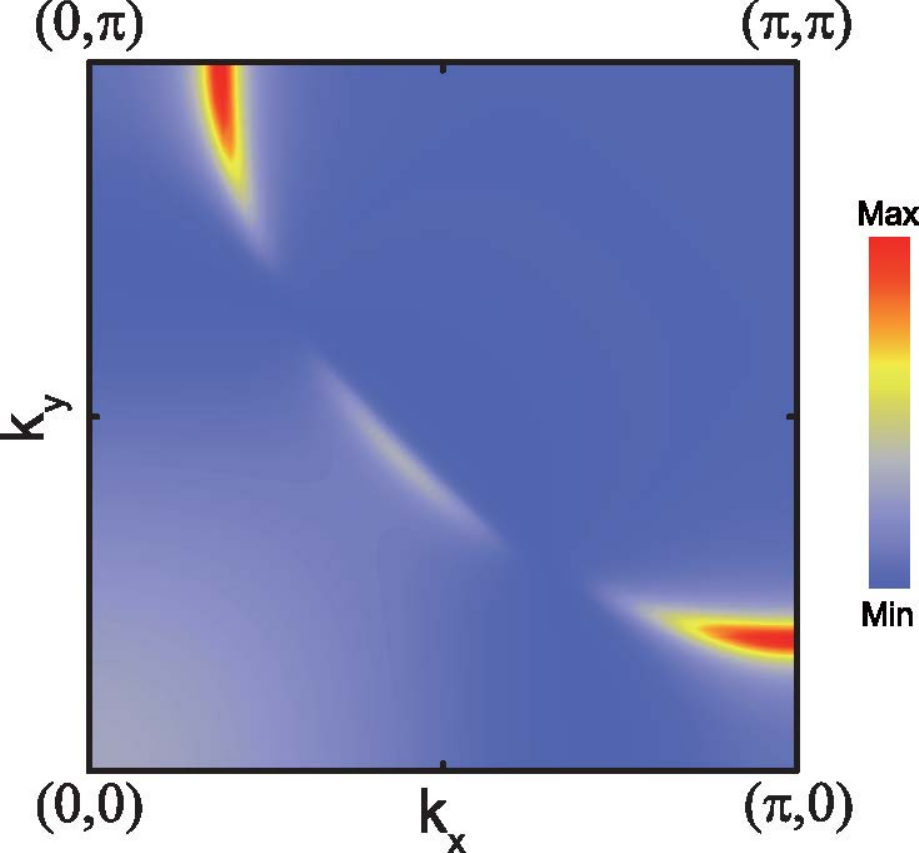}
\caption{(Color online) The map of the imaginary part of the electron self-energy at $\delta=0.12$ with $T=0.002J$ for $t/J=2.5$ and $t'/t=0.3$. \label{imaginary-part-self-energy}}
\end{figure}

The above obtained results indicate clearly that as a result of the self-consistent interplay between electrons and spin excitations in cuprate superconductors, the electron quasiparticle energies are strongly renormalized. These renormalized electron quasiparticles occupy the disconnected Fermi arcs, and then the charge-order state is driven by the Fermi-arc instability. Within the framework of the fermion-spin theory \cite{Feng15,Feng9404}, the essential physics of charge order in the pseudogap phase can be intrinsically attributed to the energy and momentum dependence of the electron self-energy $\Sigma_{1}({\bf k},\omega)$ in the particle-hole channel. This follows a fact that EFS is determined directly by,
\begin{eqnarray}\label{Fermi-energy}
\varepsilon_{\bf k}+{\rm Re}\Sigma_{1}({\bf k},0)=0,
\end{eqnarray}
in the electron spectral function $A({\bf k},\omega)$ in Eq. (\ref{full-spectral-function}) at zero energy [then the poles of the electron Green's function at zero energy], and then the lifetime of the electron quasiparticle at EFS is determined by the inverse of the imaginary part of the electron self-energy $1/|{\rm Im} \Sigma_{1}({\bf k}_{\rm F},0)|$. In Fig. \ref{imaginary-part-self-energy}, we map the intensity of the imaginary part of $|{\rm Im}\Sigma_{1}({\bf k},0)|$ at $\delta=0.12$ with $T=0.002J$. As seen from Fig. \ref{imaginary-part-self-energy}, $|{\rm Im}\Sigma_{1}({\bf k},0)|$ is strongly dependent on momentum, and has a strong angular dependence with actual maximums around the antinodal region at EFS, which leads to that the part of EFS around the antinodal region is suppressed, leaving behind disconnected segments. On the other hand, the actual minimums do not appear around the nodal points of the Fermi arcs, but locate exactly at the positions of the hot spots of the Fermi arcs. These hot spots on the Fermi arcs yield a charge-order wave vector $Q_{\rm HS}$, and therefore contribute effectively to the electron quasiparticle scattering process.

In our previous work \cite{Feng12}, it has been shown that the interaction between charge carriers directly from the kinetic energy of the $t$-$J$ model (\ref{cssham}) by the exchange of spin excitations induces the charge-carrier self-energy $\Sigma^{({\rm h})}_{1}({\bf k},\omega)$ (then the pseudogap) in the particle-hole channel and charge-carrier self-energy $\Sigma^{({\rm h})}_{2}({\bf k},\omega)$ (then the charge-carrier pairing state) in the particle-particle channel, and then the pseudogap is identified as being a region of the charge-carrier self-energy effect in the particle-hole channel in which the charge transport of cuprate superconductors can be interpreted in terms of the formation of this pseudogap \cite{Feng15,Qin14}. On the other hand, the electronic state originated from the charge-carrier state is due to the full charge-spin recombination \cite{Feng15a}, then the electron self-energies $\Sigma_{1}({\bf k},\omega)$ in the particle-hole channel and $\Sigma_{2}({\bf k},\omega)$ (then the electron Cooper pairing state) in the particle-particle channel are thought to be generated by the interaction between electrons by the exchange of spin excitations, and therefore the physics of the pseudogap state in the charge-carrier state is also true in the present electronic state \cite{Feng15a}. In this case, the electron self-energy $\Sigma_{1}({\bf k},\omega)$ in Eq. (\ref{ESE1}) also can be rewritten as \cite{Feng15a,Feng12},
\begin{eqnarray}\label{EPG}
\Sigma_{1}({\bf k},\omega) \approx {[\bar{\Delta}_{\rm PG}({\bf k})]^{2}\over \omega+\varepsilon_{0{\bf k}}},
\end{eqnarray}
where $\varepsilon_{0{\bf k}}=L^{({\rm e})}_{2}({\bf k})/L^{({\rm e})}_{1}({\bf k})$ is the energy spectrum of $\Sigma_{1}({\bf k},\omega)$, and $\bar{\Delta}_{\rm PG} ({\bf k})= L^{({\rm e})}_{2}({\bf k})/\sqrt{L^{({\rm e})}_{1}({\bf k})}$ is the pseudogap, while the functions $L^{({\rm e})}_{1}({\bf k})=-\Sigma_{\rm 1o}({\bf k},\omega=0)$ and $L^{({\rm e})}_{2}({\bf k})=\Sigma_{1}({\bf k},\omega=0)$ can be obtained directly from the electron self-energy $\Sigma_{1}({\bf k},\omega)$ in Eq. (\ref{ESE1}). As in the charge-carrier state \cite{Feng12}, the pseudogap $\bar{\Delta}_{\rm PG}({\bf k})$ is also identified as being a region of the electron self-energy effect in the particle-hole channel in which the pseudogap $\bar{\Delta}_{\rm PG}({\bf k})$ suppresses the low-energy spectral weight of the electron quasiparticle excitation spectrum. To show this point clearly, the corresponding imaginary part of $\Sigma_{1}({\bf k},\omega)$ also can be expressed explicitly in terms of Eq. (\ref{EPG}) as,
\begin{eqnarray}\label{IESE}
{\rm Im}\Sigma_{1}({\bf k},\omega) \approx  2\pi[\bar{\Delta}_{\rm PG}({\bf k})]^{2}\delta(\omega+\varepsilon_{0{\bf k}}),
\end{eqnarray}
which shows that there is an intrinsic relation between the electron quasiparticle scattering and pseudogap $\bar{\Delta}_{\rm PG}({\bf k})$, in agreement with the ARPES experiment \cite{Matt15}. The result in Eq. (\ref{IESE}) also shows that the pseudogap $\bar{\Delta}_{\rm PG}({\bf k})$ is strongly dependent on momentum, and then the product of $[\bar{\Delta}_{\rm PG}({\bf k})]^{2}$ and the delta function $\delta(\varepsilon_{0{\bf k}})$ has the same angular dependence on EFS as that of $|{\rm Im}\Sigma_{1}({\bf k},0)|$ shown in Fig. \ref{imaginary-part-self-energy}. In particular, this product of $[\bar{\Delta}_{\rm PG}({\bf k})]^{2}$ and $\delta(\varepsilon_{0{\bf k}})$ plays the same role in the suppression of EFS around the antinodal region as that of $|{\rm Im}\Sigma_{1}({\bf k},0)|$, i.e., it suppresses EFS around the antinodal region, leading to that EFS consists, not of closed contour, but only of four disconnected Fermi arcs centered around the nodal region, and then the charge-order state is driven by the Fermi-arc instability. In other words, the charge-order state is manifested within the pseudogap regime, then the Fermi arc, charge order, and pseudogap in cuprate superconductors are intimately related each other, and they have a root in common originated from the electron self-energy $\Sigma_{1}({\bf k},\omega)$ in the particle-hole channel due to the interaction between electrons by the exchange of spin excitations. Moreover, the pseudogap parameter $\bar{\Delta}_{\rm PG}$ in Eq. (\ref{EPG}) has the same doping dependence as that obtained previously from the charge-carrier self-energy in Ref. \cite{Feng12}, i.e., the magnitude of the pseudogap parameter $\bar{\Delta}_{\rm PG}$ smoothly decreases upon increasing doping \cite{Feng15a}. This doping dependence of the pseudogap parameter $\bar{\Delta}_{\rm PG}$ therefore leads to that the charge-order wave vector decreases with the increase of doping in cuprate superconductors.

Finally, it should be noted that the emergence of charge order in the pseudogap phase of cuprate superconductors has been detected with various techniques \cite{Comin15,Wu11,Chang12,Ghiringhelli12,Ghiringhelli14,Comin14,Neto14,Hashimoto15,Comin15a} including ARPES, scanning tunneling (STM), resonant X-ray scattering (REXS), and other methods. In particular, the experimental results from the combined ARPES, STM, and REXS indicate a charge order that appears consistently in surface and bulk, and in momentum and real space in cuprate superconductors \cite{Comin14}. In this case, the key questions surrounding charge order and its relevance to superconductivity are raised \cite{Comin15,Wu11,Chang12,Ghiringhelli12,Ghiringhelli14,Comin14,Neto14,Hashimoto15,Comin15a}: (a) Are the unusual properties observed from STM, REXS, and other experimental methods the result of the emergence of charge order? (b) Do superconductivity and charge order coexist? (c) Is charge order a necessary ingredient for superconductivity? These and the related issues are under investigation now.

\section{Conclusions}\label{conclusions}

In conclusion, within the framework of the fermion-spin theory, we have studied the nature of charge order and its evolution with doping in cuprate superconductors in the pseudogap phase by taking into account the electron self-energy (then the pseudogap) effect. Our result shows that the antinodal region of EFS is suppressed by the electron self-energy, and then the low-energy electron excitations occupy the disconnected Fermi arcs located around the nodal region. The charge-order state is driven by the Fermi-arc instability, with a characteristic wave vector corresponding to the hot spots of the Fermi arcs rather than the antinodal nesting vector. In particular, the Fermi arc in the underdoped regime increases in length with doping, while the charge-order wave vector reduces almost linearity with increasing doping. Our theory also indicates that there is a common origin for the Fermi arc, charge order, and pseudogap, and they are a natural consequence of the strong electron correlation.

In cuprate superconductors, mangy different kinds of ordered states, such as superconductivity, charge order (then pseudogap), and antiferromagnetism, occur on comparable temperature scales. When they compete, they do so on an almost equal footing \cite{Chang12}. Within the framework of the kinetic-energy-driven SC mechanism, we have shown that the quasiparticle coherence, related directly to the pseudogap, antagonizes superconductivity, and then $T_{\rm c}$ is depressed to low temperatures \cite{Feng15,Feng0306,Feng12}. Since there is an intimate connection between the pseudogap and charge order, charge order has a competitive role in engendering superconductivity. This kinetic-energy-driven superconductivity also indicates that cuprate superconductors have two ground-states: a pseudogap (then charge-order) state above $T_{\rm c}$, which involves the electron correlation from the particle-hole channel, versus superconductivity below $T_{\rm c}$, which coexists with the charge-order (then the pseudogap) state, and involves the electron correlation from the particle-particle channel, and then the phase diagram as a whole is driven by the spin-excitation mediated interaction.

\section*{Acknowledgements}

The authors would like to thank L\"ulin Kuang, Xixiao Ma, Ling Qin, and Yu Lan for helpful discussions. SF and DG are supported by the funds from the Ministry of Science and Technology of China under Grant No. 2012CB821403, and the National Natural Science Foundation of China (NSFC) under Grant Nos. 11274044 and 11574032, and HZ is supported by NSFC under Grant No. 11447144.

\end{document}